\documentclass[aip,graphicx]{revtex4-1}
\usepackage{graphics}
\usepackage{bm}
\usepackage{subfigure}
\usepackage{epsfig}
\usepackage[bookmarks,colorlinks,citecolor=blue,linkcolor=blue]{hyperref}
\usepackage{natbib}
\usepackage{siunitx}
\usepackage{soul}
\usepackage{array}
\usepackage{multirow}

\DeclareSIUnit\Molar{\textsc{m}}
\newcommand{\beginsupplement}{%
        \setcounter{table}{0}
        \renewcommand{\thetable}{S\arabic{table}}%
        \setcounter{figure}{0}
        \renewcommand{\thefigure}{S\arabic{figure}}%
     }
\usepackage{xcolor}
\usepackage[normalem]{ulem}

\draft 

\begin{document}


\title{Topological insights into dense frictional suspension rheology: Third order loops drive discontinuous shear thickening} 
\author{Alessandro D'Amico}
\affiliation{Department of Chemical Engineering, Case Western Reserve University, Cleveland, OH, 10040, USA}
\author{Sidong Tu}
\affiliation{Department of Macrocalmolecular Science and Engineering, Case Western Reserve University, Cleveland, OH, 10040, USA}
\author{Abhinendra Singh}
\email{abhinendra.singh@case.edu}
\affiliation{Department of Macromolecular Science and Engineering, Case Western Reserve University, Cleveland, OH, 10040, USA}

\date{\today}

\begin{abstract}
Dense suspensions exhibit significant viscosity changes under external deformation, a phenomenon known as shear thickening. Recent studies have identified a stress-induced transition from lubricated, unconstrained interactions to frictional contacts, which play a crucial role in shear thickening. This work investigates the rheological behavior and contact network evolution during continuous and discontinuous shear thickening (CST and DST) in two-dimensional simulations. We find that at low stress, during weak thickening, the frictional contact network is composed of quasilinear chains along the compression axis. With increasing stress, the contact network becomes more isotropic, and forms loop-like structures. We show that third-order loops within the frictional contact network are key to this behavior. Our findings revealed a strong correlation between the number of edges in the third-order loops and the viscosity of the suspension. Notably, this relationship remains independent of the packing fraction, applied stress, and interparticle friction, highlighting the fundamental role of the mesoscale network topology in governing macroscopic rheology.
\end{abstract}

\pacs{}

\maketitle 

\section{Introduction}\label{Intro}
Suspensions of small particles in a viscous liquid, a prototypical class of disordered amorphous materials, often exhibit enhanced resistance to flow (i.e., increased viscosity) under large deformation, a phenomenon known as shear thickening~\cite{Brown_2014, Morris_2020, Singh_2023}.
When the particle concentration, or packing fraction $\phi$, approaches but remains below the jamming limit $\phi_J^\mu$, the shear thickening behavior becomes abrupt, occurring over a narrow range of shear rates. This phenomenon is referred to as ``discontinuous shear thickening'' (DST)\cite{Wyart_2014, Mari_2014, Ness_2016, Singh_2018, Singh_2020, Morris_2020, Singh_2022}.
Extensive experimental studies have demonstrated that DST occurs under a wide range of conditions, including the variations in particle size\cite{Maranzano_2001, Guy_2015}, stabilization mechanisms~\cite{Guy_2015}, and interfacial chemistry~\cite{Maranzano_2001a, James_2018, Laun_1984}. This generality suggests a unifying physical mechanism underlying the DST across diverse systems.
%
At packing fractions $\phi > \phi_J^\mu$, the suspension transitions into a solid-like shear-jammed (SJ) state under large deformation~\cite{Peters_2016, Barik_2022}. This solid-like SJ state, formed under shear, is distinct from the unyielded viscoelastic soft solids typically studied in the context of colloidal gels~\cite{Zaccarelli_2007, Vinutha_2023,Mangal_2023,Delgado_2016}. Unlike colloidal gels, which can sustain a load without continuous deformation, SJ solids rely on a dynamic, shear-induced contact network to support stress~\cite{Morris_2020, Bi_2011, Cates_1998a, Seto_2019}.
The SJ state is fragile: the load is maintained only when the material is steadily sheared—any change in the direction or magnitude of loading results in temporary material failure. However, over time, steady shearing in the new direction can re-establish the contact network, forming a shear-jammed solid aligned with the new flow direction~\cite{Seto_2019, Cates_1998a}.

Recent studies suggest that the breakdown of lubrication films, leading to the formation of frictional contacts between particles as the applied stress exceeds the onset stress $\sigma_0$, is the key mechanism driving the shear thickening behavior~\cite{Mari_2014, Wyart_2014}. This phenomenon represents a stress-activated transition from an unconstrained, lubricated ``frictionless'' state to a constrained state~\cite{Guy_2018, Singh_2020, Singh_2022,Singh_2024}. These constraints, which originate from frictional contacts~\cite{Hsu_2018, Clavaud_2017, Seto_2013a, Mari_2014, Singh_2018, Singh_2020} or a combination of hydrodynamic forces and surface asperities~\cite{Jamali_2019}, significantly hinder the relative motion between particle pairs.
Constraint-based mean-field models have been particularly successful in capturing the steady-state, strain-averaged flow behavior of dense suspensions and effectively describing data from both experiments and simulations~\cite{Guy_2015, Ness_2016, Singh_2018, Han_2019, Ramaswamy_2023}. These models predict a relationship between stress-induced frictional shear thickening and shear jamming, suggesting that strong shear thickening, or discontinuous shear thickening (DST), serves as a precursor to shear jamming~\cite{Singh_2018, Peters_2016}.

Although insightful, the aforementioned mean-field approaches do not account for the topological and geometrical characteristics of the mesoscale force and contact networks. These networks play a critical role in resisting external deformation, leading to enhanced viscosity and normal stresses.
Recent studies have investigated the relationship between the underlying force and contact network structure and the rheology of both shear-thickening suspensions~\cite{Boromand_2018, Thomas_2018, Sedes_2022, Gameiro_2020, Edens_2019, Edens_2021, Naald_2024, Nabizadeh_2022,Goyal_2022} and colloidal gels~\cite{Nabizadeh_2024, Mangal_2023}. 
%
%
Based on the observation in prior studies~\cite{Mari_2014} showing the changes in pair-correlation functions in real space, showing minor changes across the DST transitions; Thomas et al.~\cite{Thomas_2018, thomas2020investigating} suggested that the network space (orthogonal to the particle space) plays a central role in understanding the statistical physics of frictional suspensions.
Sedes \textit{et al.} recently employed $K$-core analysis to examine the particle clusters within a percolated contact network~\cite{Sedes_2022}. In addition, Nabizadeh \textit{et al.}~\cite{Nabizadeh_2022} applied community detection methods to reveal distinct interactions between clusters at packing fractions where suspensions transition from continuous shear thickening (CST) to discontinuous shear thickening (DST).
Network science tools were initially applied to particulate systems to understand shear banding, jamming-unjamming, and related phenomena in dry granular materials, as reviewed by Papadopoulos \textit{et al.}~\cite{Papadopoulos_2018}.

Previous studies on dry granular materials have highlighted a connection between the onset of rigidity, mechanical equilibrium, and characteristics of the contact network, particularly the presence of loops~\cite{Arevalo_2010, Smart_2008}. 
This concept gained prominence through the observation of odd circuits in granular systems, referred to as R-loops~\cite{Blackett01121979}, which are used to explain the origin of rigidity in granular packings~\cite{RIVIER20064505}. These contact loops are known to contribute to the stability of granular materials~\cite{RIVIER20064505, mankoc2007flowrategranularmaterials} and have provided valuable insights into the transition from fluid- to solid-like behavior~\cite{Arevalo_2010}, as well as the stress response~\cite{Smart_2008} in granular packings. It is important to note that the term ``cycle'' ass also been used to describe such circuits~\cite{Tordesillas_2010, Giusti_2016}. In this study, we adopt the term ``loop'' to topologically characterize the force network and, thus, maintain consistency with the existing literature~\cite{Arevalo_2010, Smart_2008, Tordesillas_2010}.

In this study, we applied the loop-based network approach to a new context. Rather than analyzing the structure of loops in a static configuration, we examine the loop structure within a dynamically evolving network of frictional contacts in a dense suspension subjected to simple shear flow.
This study aims to extract mesoscale features from the dynamical frictional contact network and correlate these features with viscosity, which depends on the packing fraction $\phi$, applied stress $\sigma$, and particle friction $\mu$.
We demonstrate that the number of third-order loops, the smallest minimally rigid structures (in Lamans' theorem~\cite{Laman_1970}), correlate strongly with pronounced shear thickening behavior. Specifically, we show that the viscosity $\eta_r$ increases with the number of third-order loop edges and appears to diverge as the number of these loops reaches a critical threshold.
Moreover, this correlation is largely insensitive to particle friction, making third-order loop edges a valuable order parameter for identifying transitions in z material state.

\section{Methodology}\label{Methods}
Our strategy for relating rheology with the mesoscale characterization proceeds in three parts: first, the physical data of particle position and interactions are mapped to a set of nodes and edges relevant to network science; second, we analyze the topological features of this network as the system is sheared; and third, we extract relevant features and study how our parameter space ($\phi,\sigma,\mu$) affects them. Finally, we correlate the flow behavior (viscosity) with the topological features of the frictional network.

\subsection{Simulating dense suspensions}
Although real-world dense suspensions are three-dimensional, related prior works have shown the flow behavior for 2D and 3D to be similar if the packing fraction $\phi$ is appropriately scaled~\cite{Nabizadeh_2022, Gameiro_2020}.
Given that loops are two-dimensional planar concepts, we performed two-dimensional simulations (a monolayer of spheres) in this study.
 The simulation scheme used here is, in principle, similar to that in the prior work by Mari \textit{et al.}~\cite{Mari_2014, Mari_2015} with the only difference being that we simulate a two-dimensional monolayer of spheres. In this study, a series of stress-controlled simulations are performed to analyze the frictional contact network in suspension close to but below the frictional jamming point $\phi_J^\mu$.
Our simulation scheme (LF-DEM) includes hydrodynamics interactions and repulsive contact forces that incorporate static friction. Hydrodynamic interactions include the one-body Stokes drag and two-body lubrication forces (LF). The contacts between two particles are modeled using the discrete element method (DEM) from dry granular materials~\cite{Seto_2013a, Mari_2014, Morris_2020, Singh_2020}.
Ignoring the full hydrodynamics is strategic here with twofold aims: (i) We base ourselves on prior works by Singh et al.~\cite{Singh_2020,Singh_2022} where a quantitative comparison has been reported considering lubrication interactions only; and (ii) including the full long-ranged hydrodynamic interactions makes the computation intractable~\cite{Ball_1995, Morris_2020}. Shear is implemented using Lees-Edwards boundary condition under fixed shear stress with constant volume. To avoid crystallization, we simulate a particle mixture with a size ratio of 1.4 mixed in an equal volume.

Owing to the Stokes flow, that is, the inertialess limit, the particles obey the overdamped equation of motion 
\begin{equation}
0 = \vec{F}_{\mathrm{H}}(\vec{X},\vec{U}) + \vec{F}_{\mathrm{C}}(\vec{X})~,
\end{equation}
where $\vec{X}$ and $\vec{U}$ denote the particle positions and velocities, respectively. Here, $\vec{F}_{\mathrm{H}}$, $\vec{F}_{\mathrm{C}}$ denote the hydrodynamic and contact forces, respectively. At every step, the overall stress in the suspension is fixed and is given by the sum of the hydrodynamic $\sigma_\mathrm{H}$ and contact $\sigma_\mathrm{C}$ contributions as follows:
\begin{equation}
    \sigma = \Sigma_{xy} = \dot{\gamma}\eta_0 \biggl( 1+\frac{5}{2}\phi \biggr) + \dot{\gamma}\eta_{\mathrm H} + \sigma_{\mathrm C}~.
\end{equation}
Here, $\eta_0$ is the suspending liquid viscosity, 
$\eta_{\mathrm H} = V^{-1}\bigl\{(\bm{R}_{\mathrm {SE}}-\bm{R}_{\mathrm {SU}}\cdot\bm{R}_{\mathrm {FU}}^{-1}\cdot\bm{R}_{\mathrm {FE}}):\hat{\bm{E}^\infty}\bigr\}_{xy}$, 
$\sigma_{\mathrm C} = V^{-1}\bigl\{\bm{X}\bm{F_C}-\bm{R}_{\mathrm {SU}}\cdot\bm{R}_{\mathrm {FU}}^{-1}\cdot\bm{F_C}\bigr\}_{xy}$, 
$R_{\mathrm {SU}}$ and $R_{\mathrm {SE}}$ are the resistance matrices used to calculate the lubrication stress~\cite{Mari_2014, Jeffrey_1992}, ${E^\infty}$ denotes the rate-of-strain tensor,  and $V$ is the volume of the simulation box.
$R_{\mathrm {FU}}$ and $R_{\mathrm {FE}}$ denote the position dependent resistance matrices containing the ``squeeze,'' ``shear,'' and ``pump'' modes of pairwise lubrication along with one-body Stokes drag.
Imposing constant shear stress leads to a time-dependent shear rate $\dot{\gamma}(t)$, which is used to calculate the fluctuating viscosity $\eta_r (t) = \sigma/\dot{\gamma}(t)$. 

Lubrication is regularized, allowing the particles to come into contact as the overlap ($\delta^\text{(i,j)} = a_i + a_j - |\vec{r_i} - \vec{r_j}|$) becomes positive~\cite{Ball_1995, Mari_2014}. Here, $a_i$ and $a_j$ are the particle radii, and $\vec{r_i}$ and $\vec{r_j}$ denote the position vectors of the center of particles $i$ and $j$, respectively.
In our simulations, we consider both normal and tangential frictional forces, i.e.,  $\vec{F}_{\mathrm{C}} = \vec{F}_{\mathrm{C}}^N+\vec{F}_{\mathrm{C}}^T$; tangential and normal contact forces satisfy $|\vec{F}_{\mathrm{C}}^t \le \mu |\vec{F}_{\mathrm{C}}^n|$, with $\mu$ being the static friction coefficient.
We followed the approach of Cundall and Strack \cite{Cundall_1979}
and the algorithm by Luding~\cite{Luding_2008} to model the contact between particles. 
Linear springs in the normal $k_n$ and tangential $k_t$ directions are used to model the contacts between particles. Both $k_n$ and $k_t$ are tuned to allow the maximum overlap to not exceed 3\% of the particle size to maintain rigid particle approximation~\cite{Singh_2015}.

Finally, we employ the Critical Load Model (CLM) to introduce the rate dependence, where ${F}_{\mathrm{C}}^N \ge F_0$ is required to activate the Coulombic static friction $\mu$ between the particles. This critical force $F_0$ leads to a characteristic stress scale $\sigma_0 = F_0/6\pi a^2$, such that for $\sigma \ll \sigma_0$ the particle interactions are lubricated, whereas for $\sigma \gg \sigma_0$ almost all contacts are frictional. For further details on the simulation scheme for modeling the hydrodynamics and contact forces, see~\cite{Mari_2014, Mari_2015, Singh_2018, Singh_2020}.
In the remainder of this work, quantities are reported in dimensionless form in terms of $\eta_r = \eta/\eta_0$, $\tilde{\sigma} = \sigma/\sigma_0$ and $\dot{\gamma}/\dot{\gamma}_0$ with $\dot{\gamma}_0 \equiv F_{\rm 0}/{6\pi \eta_0 a^2}$.
%

\begin{figure}[ht!]
    \centering
    \includegraphics[width=1\textwidth]{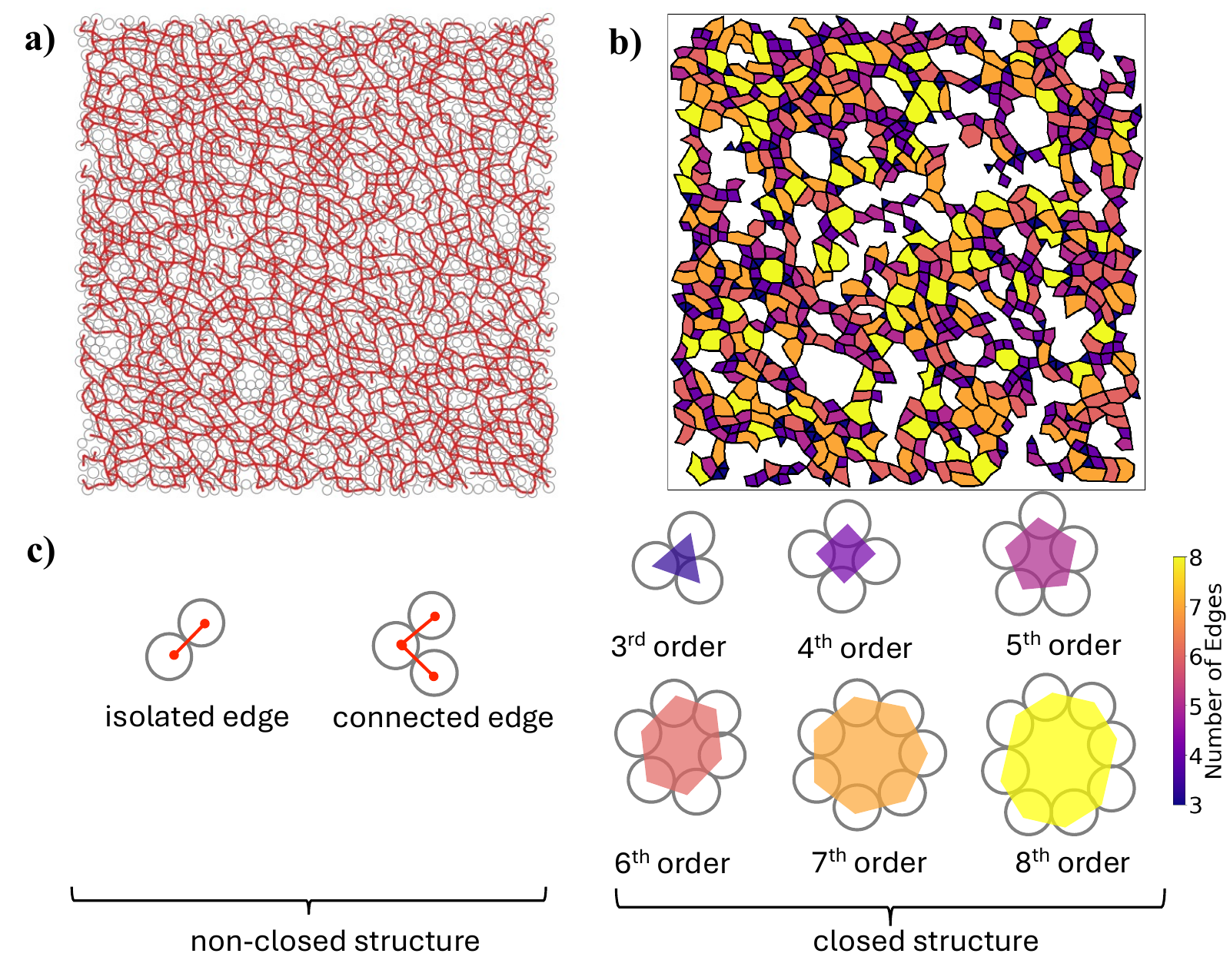}
    \caption{\textbf{Decomposition of the frictional contact network into loops}. (a) Simulation snapshot with frictional contact, where normal force exceeds the critical threshold, shown in red. (b) Frictional contact network in \textbf{a}, decomposed into loops.  The loop shapes are filled according to the number of edges where warmer colors (yellow) indicate a larger loop order and cooler colors (purple) indicate a smaller loop order, as described in (c).}
    \label{fig1-def}
\end{figure}

\subsection{Loop finding algorithm}
This work focuses on the frictional contact network (FCN), which plays a crucial role in determining viscosity at high packing fractions and plays a key factor in discontinuous shear thickening (DST) and shear jamming (SJ) behaviors~\cite{Mari_2014, Singh_2020, Seto_2013a, Ness_2016, Singh_2018, Hsu_2018, Clavaud_2017}. Figure~\ref{fig1-def}a presents a typical FCN, as obtained from the simulations conducted in this study, where the red bonds represent the frictional contacts between the particles. 

To analyze the data, we converted the physical space into a graph $G$ for each snapshot (Fig.~\ref{fig1-def}a). In graph theory, each particle is represented as a node, and an edge exists between the two nodes only if the particles are in frictional contact. By definition, the contact network is undirected and unweighted~\cite{Papadopoulos_2018}. Note that we focus on the frictional contact network rather than the frictional force network, which is weighted (except for the analysis presented in Fig.~S1). We deal with a dynamic system, as the suspension is constantly under shear, meaning that contacts continually form, break, and reform. As a result, the network (or graph) is also dynamic and is analyzed snapshot by snapshot.
A fundamental local property of any graph $G$ is the node \textit{degree}, which, in an unweighted network, is simply the number of edges $m$ attached to a node. The mean degree for $N$ nodes is $\langle k \rangle = \frac{2m}{N}$, often referred to as the coordination number $Z$. In this study, we use $Z_\mu$ to denote the frictional coordination number or the number of frictional contacts per particle. In graph theory, a \textit{walk} is a traversal from one node to another along the edges of a network. A simple \textit{path} is a walk that does not revisit the same node or edge. A \textit{closed path} is a special case where the walk returns to the starting node, which we refer to as a loop (or cycle, in some cases). A loop of order $l$ is a closed path traversing $l$ edges. In network science, the loops (or cycles) considered here are simple cycles, meaning no node or edge is revisited except for the starting and ending nodes. Thus, the smallest loop considered is a third-order loop. We refer to the non-closed paths as isolated edges or connected edges, depending on the number of nodes involved (Fig.~\ref{fig1-def})c.

\begin{figure}[ht!]
    \centering
    \includegraphics[width=0.35\textwidth]{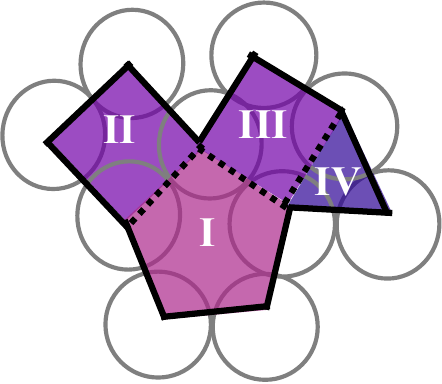}
    \caption{\textbf{Schematic of loop edge attribution.} An example demonstrating the edge attribution while loops are counted. The solid and dotted lines represent the individual and shared edges, respectively. 
The numbers (I, II, III, and IV) denote and represent loops: fifth-order loop (I), fourth-order loop (II and III), and third-order loop (IV). The color coding for the loop is the same as in Fig.~\ref{fig1-def}. }
    \label{fig2-exp}
\end{figure}

In this study, we restricted our analysis to loops of order $l \le 8$, with the rationale for this limitation discussed at the end of the section. This approach is equivalent to identifying all loops from the third- to eighth-order. In our analysis, we prioritize the smaller-order loops. Specifically, if at least one edge in a loop of order $l$ does not belong to a loop of smaller order, all edges are initially assigned to the loop of order $l$.
During the assignment of edges to a specific loop of order $l$, we encounter two possible scenarios: an edge may either belong exclusively to a particular loop, referred to as an ``individual edge,'' or it may be shared between two loops, referred to as a ``shared edge'' (Ref. Fig.~\ref{fig2-exp}). Based on this classification, we assign attributes to the edges: a shared edge is assigned an attribute of 0.5, whereas an individual edge is assigned an attribute of 1.
It is important to note that in a planar two-dimensional system, an edge can be shared by two loops, whereas in a three-dimensional system, the number of shared loops may be higher.
%
An attribute of 0.5 in the case of shared edges ensures that while summing all the edges belonging to loops of different orders $3\le l \le 8$, we recover the degree of the network.
Figure~\ref{fig2-exp} illustrates this concept. The fifth-order loop (I) has three individual edges (solid lines) and two shared edges (dotted lines). We summarize the number of edges based on the edge attribution and divide the sum by the size of loops (five for the fifth order loop), which leads to 0.8 fifth order loops. Similarly, in this example, we have a 1.625 fourth-order loops and a 0.833 third-order loops.

%
%

Once all edges belonging to loops of orders 3-8 are found, the remaining edges are considered \textit{connected edges} if at least one of the two nodes constituting that edge has at least one other frictional contact (Fig.~\ref{fig1-def}c). If neither node has another frictional contact, meaning that both nodes have a degree of one, then the edge is considered an \textit{isolated edge}. 


%

Our approach differs from previous methods, particularly the one proposed by Smart \& Ottino~\cite{Smart_2008}, in two key ways: edge-centric approach and edge sharing. Smart and Ottino assert that every angle defined as two edges sharing a common vertex (AKA node) —must belong to only one loop, with the angle assigned to the smallest-order loop. By contrast, the algorithm presented in this study is based on edge allocation rather than angle allocation. When we tested alternative edge-sharing schemes, we found that the results for third-order loops were minimally affected, leading to the same conclusions. Ultimately, the edge-sharing scheme we presented was selected for its more concrete physical interpretation.

Finally, we provide the rationale for our choice of limiting the analysis to loops of order $l \le 8$: (i) when calculating the average stress carried by each loop, we observe that loops up to the eighth-order typically carry the contact stress borne by the packing, while the contributions of higher-order loops are not significantly different from those of connected edges (Fig. S1 in SI); (ii) In the context of subgraph isomorphism, loops with $l>8$ can exhibit numerous isomorphic cases, complicating the analysis; and (iii) the computational cost of detecting higher-order loops increases significantly, eventually becoming untractable. For these reasons, we restrict our focus to eighth-order loops in this study.

Simulations for a given condition $(\phi,\tilde{\sigma},\mu)$ were performed over 40 strain units with a strain step of $\Delta(\gamma)=0.1$. This sampling frequency ensures that each snapshot is statistically independent, meaning that each simulation snapshot samples a distinct ensemble of the frictional network (and thus loops). This independence allows for an accurate correlation between the network topology and strain-averaged rheological quantities, such as viscosity.
At each step, we converted the snapshot to graph $G$ and applied the loop-detection algorithm to extract information about isolated edges, connected edges, and the number of loops $n_l$ for $3\le l \le 8$.

\section{Results}\label{Results}
Figure~\ref{fig1-def}a shows a typical frictional contact network obtained from our simulation for the conditions $\{\phi,\tilde{\sigma},\mu\} = \{0.78, 100,1\}$. The line segments connecting the centers of the two particles in frictional contacts ($f_n>F_0$) are represented by red bonds. Applying our loop detection algorithm, we decompose this network of frictional contacts into loops of varying orders $l$, illustrated as filled polygons with different numbers of sides, as shown in Fig.~\ref{fig1-def}b. The polygons are color-coded to indicate the loops of different sides, with the number of sides corresponding to ``loop order $l$.''
Visual inspection revealed that our system consists of loops with orders ranging $3 \le l \le 8$. Based on previous studies on dry granular materials, we focus primarily on low-order loops (3rd and 4th)~\cite{Arevalo_2010, Tordesillas_2010, Smart_2008}. 
Consistent with previous studies~\cite{Smart_2008}, we find that third and fourth order loops dominate the network and the number of loops $n_l$ decrease with loop order $l$ (Fig. S2 in SI).  
%

%
%


\paragraph*{Effect of packing density.}

\begin{figure}[ht!]
    \centering
    \includegraphics[width=1\textwidth]{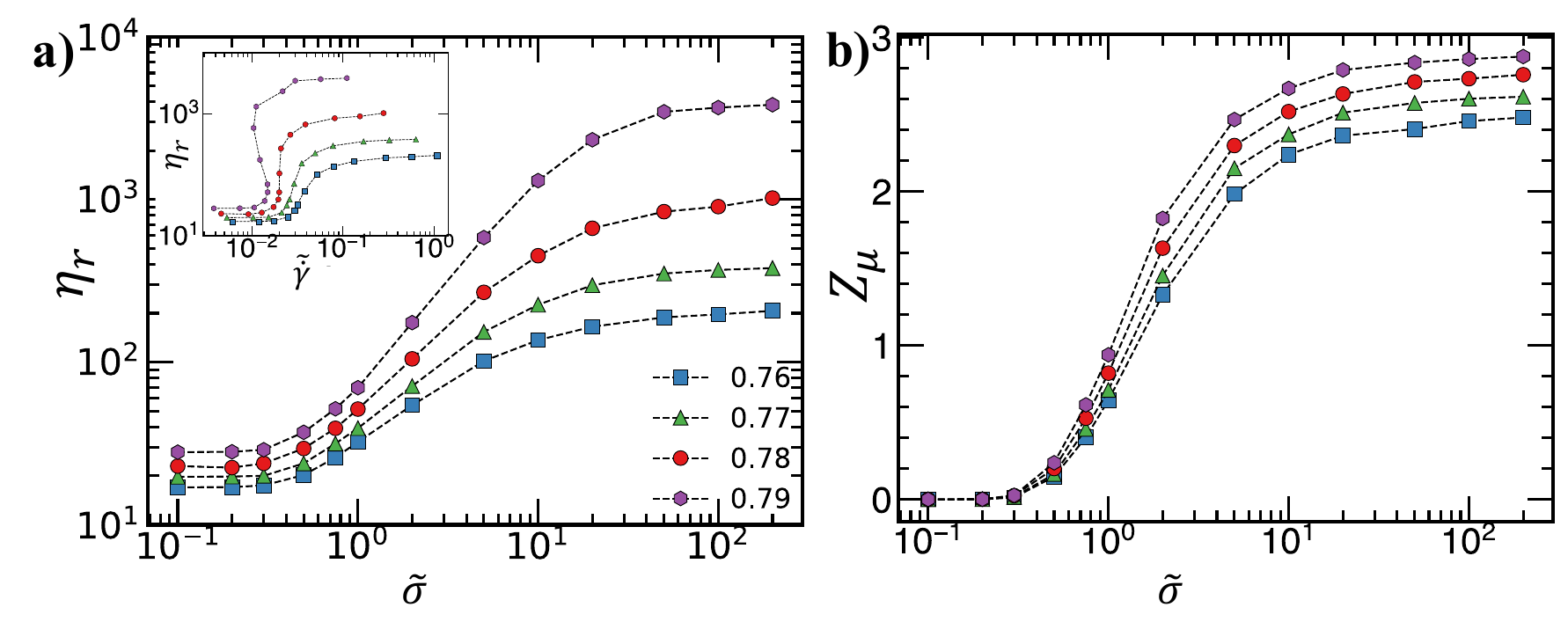}
    \caption{\textbf{Rheology and coordination number.} (a) Relative viscosity $\eta_r$  and (b) average coordination number 
    $Z_{\mu}$ as a function of scaled stress $\tilde{\sigma}$ for different packing fractions $\phi$ for simulations with $\mu$=1. With increasing $\phi$, both $\eta_r$ and $Z_\mu$ increase and saturate in the shear-thickened state. For (a), relative viscosity $\eta_r$ as a function of shear rate $\tilde{\dot{\gamma}}$ is shown in the inset. In 2D simulations performed here, DST occurs at roughly $\phi=0.78$.}
    \label{fig4-rheofixmu}
\end{figure}

We begin by examining the parameter space $\{\phi, \tilde{\sigma}\}$ for $\mu=1$ in the vicinity of $\phi_J^{\mu=1}$ to explore the effects of increasing the packing fraction $\phi$ and the applied stress $\sigma$. Figure~\ref{fig4-rheofixmu} presents a family of steady-state flow curves $\eta_r(\tilde{\sigma})$ and the coordination number $Z_\mu(\tilde{\sigma})$ for various packing fractions $\phi$.
Figure~\ref{fig4-rheofixmu}a shows the canonical shear-thickening behavior observed in previous experimental~\cite{Cwalina_2014, Guy_2015, Guy_2018, Royer_2016} and simulation~\cite{Mari_2014, Singh_2018, Ness_2016} studies in three dimensions, showing a stress-driven transition between two rate-independent plateaus. Shear thickening behavior becomes more pronounced with increases with $\phi$; at $\phi=0.76$ and 0.77, continuous shear thickening (CST) is observed, while discontinuous shear thickening (DST) occurs at $\phi=0.78$ and 0.79, as indicated by the non-monotonic flow curve $\eta_r(\tilde{\dot{\gamma}})$ (inset of ~\ref{fig4-rheofixmu}a). 
Furthermore, Fig.~\ref{fig4-rheofixmu}b shows that this transition is driven by the activation of the frictional contacts. Initially, $Z_\mu$ is zero at low stresses, and it increases once the stress exceeds a threshold of $\tilde{\sigma}=0.3$, eventually reaching a saturation point in the shear-thickened state. Notably, $Z_\mu$ in the shear-thickened state increases with $\phi$, which is consistent with the literature~\cite{Nabizadeh_2022}.

\begin{figure}[ht!]
    \centering
    \includegraphics[width=1\textwidth]{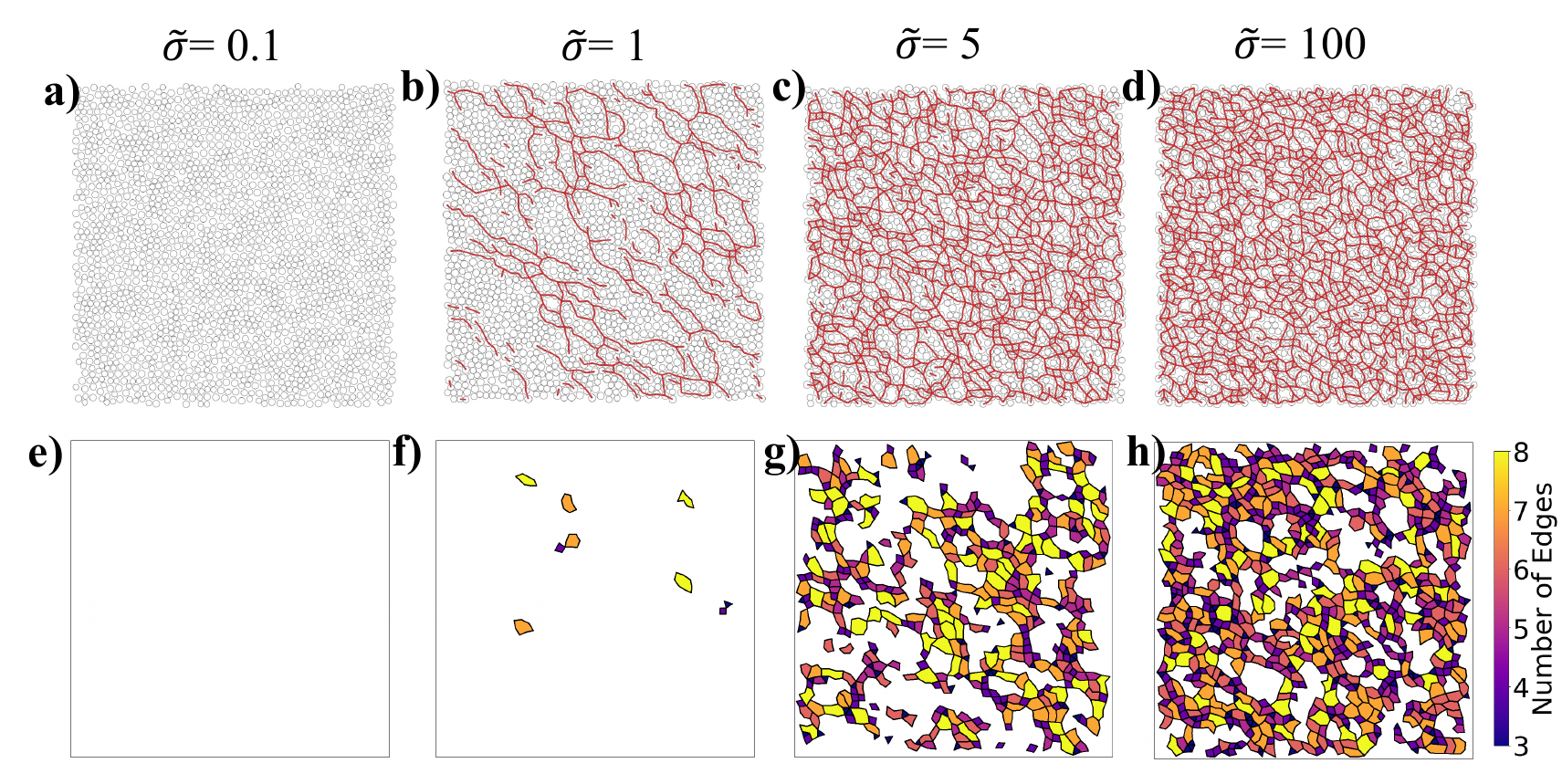}
    \caption{\textbf{Frictional contact network and corresponding loop structure.}
    Simulation snapshot with frictional contact shown in red at four stress levels ($\tilde{\sigma}$): 0.1 (a), 1 (b), 5 (c) and 100 (d),  for $\phi=0.78$ and $\mu=1$. The loops of order $l$ are displayed using polygons with $l$ sides at stress levels ($\tilde{\sigma}$): 0.1 (e), 1 (f), 5 (g), and 100 (h). The color coding for the loop is the same as that in Fig.~\ref{fig1-def}.}
    \label{fig3-timeevo}
\end{figure}

To gain mesoscale insight into DST, we analyze representative snapshots of frictional forces and their corresponding loop order-based decomposition as shown in the top and bottom panels of Fig.~\ref{fig3-timeevo}, respectively.
In Fig.~\ref{fig3-timeevo}, we show the stress-induced frictional contact networks at various stages: the lubricated state ($\tilde{\sigma}=0.1$), onset of shear thickening (ST) ($\tilde{\sigma}=1$), DST regime ($\tilde{\sigma}=5$), and deep into the shear-thickened state ($\tilde{\sigma}=100$). 
At low stress $\tilde{\sigma}=0.1$, the suspension is in the lubricated state, with no frictional contacts. At $\tilde{\sigma}=1$, the force chains emerge as roughly linear structures along the compression axis $y=-x$ of the simple shear flow $u_x=\dot{\gamma}y$ resulting in only a few fleeting loops that usually appear at the meeting or splitting point of the force chains.
At $\tilde{\sigma}=5$, frictional contacts form loops in the DST regime. Finally, at $\tilde{\sigma}=100$, the suspension reaches a fully shear-thickened state, and the network exhibits more loops than in the DST regime. 
Integrating these visual observations with the rheological data from Fig.~\ref{fig4-rheofixmu}, we observe a clear correlation between shear thickening and network topology.
Supplementary figures S3-S4 along with Video 1 show the evolution of interparticle forces and loops with packing fraction and strain for the systems shown in Fig.~\ref{fig3-timeevo} b-d.

\begin{figure}[ht!]
    \centering
    \includegraphics[width=1\textwidth]{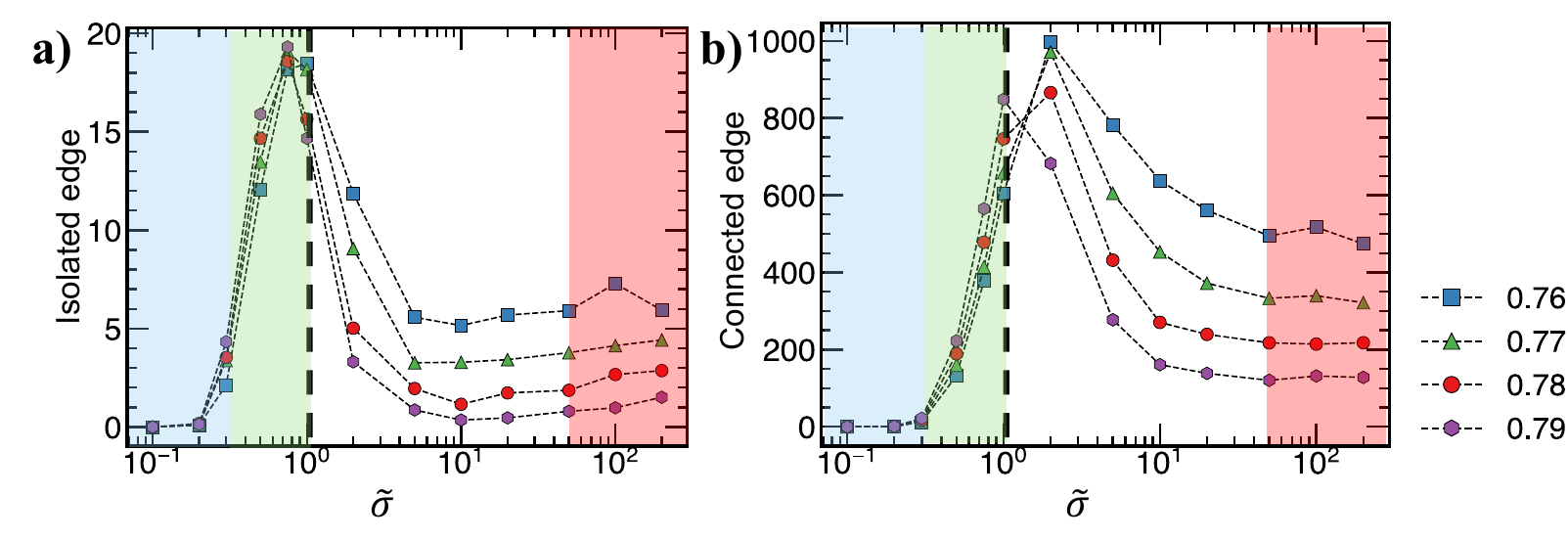}
    \caption{\textbf{Evolution of non-closed components.}
    The number of (a) isolated edges and (b) connected edges as a function of scaled stress $\tilde{\sigma}$  at $\mu$=1 for different packing fractions $\phi$. The blue blocks indicate the frictionless state where frictional contacts are absent; the green blocks indicate the stress regime where only linear force network (isolated and connected edges) are formed; the dashed lines indicate the point $\tilde{\sigma}$ =1 where higher order loop start to appear; and the red blocks show the shear-thickened stage.
 }
    \label{fig5-evon12}
\end{figure}

To quantify these visual observations, snapshots are processed using the loop-detection algorithm to detect loops and extract the number of loops $n_l$ ($3 \le l\le 8$) across different packing fractions $\phi$ and stresses $\tilde{\sigma}$.
First, we present the evolution of the non-closed structures, namely, isolated and connected edges, Figs.~\ref{fig5-evon12} a and b, respectively. 
Both quantities start at zero in the frictionless state (blue shaded region), increase as the suspension undergoes shear thickening (green shaded region), and decrease once the stress surpasses $\tilde{\sigma}>1$.
Note that both these quantities are insensitive to $\phi$ as the suspension shear thickens but show a decrease with $\phi$. At intermediate stress values $0.3<\tilde{\sigma}<1$, most of the force chains are linear structures aligned along the compressive axis, increasing isolated and connected edges. 
However, as the number of frictional contacts increases further with stress $\tilde{\sigma} \ge 2$, disconnected force chains are rare, giving way to loops that become more probable (Ref.~Fig.~\ref{fig3-timeevo}). 

\begin{figure}[ht!]
    \centering
    \includegraphics[width=1\textwidth]{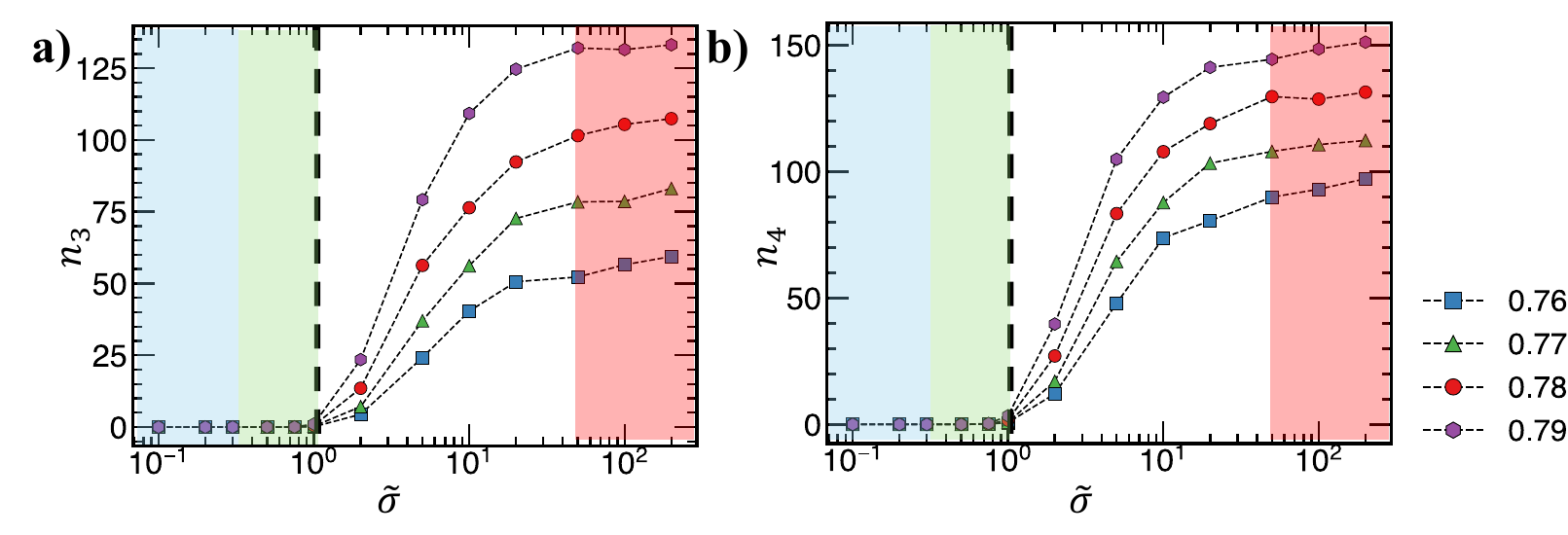}
    \caption{\textbf{Evolution of loops.} Number of (a) third order ($n_3$) and (b) forth order ($n_4$) loops as a function of scaled stress $\tilde{\sigma}$ at $\mu$=1 for different packing fractions $\phi$. The color coding for the loop is the same as in Fig.~\ref{fig5-evon12}.
 }
    \label{fig6-evon34}
\end{figure}

Next, Fig.~\ref{fig6-evon34} shows the evolution of the two smallest closed components, namely the third- and fourth-order loops, as a function of the scaled stress.
Loops are absent for stresses $\tilde{\sigma}<1$, with both $n_3$ and $n_4$ remaining zero, and they begin to increase with stress thereafter. Notably, the stress at which the loops appear coincides with the stress at which the isolated and connected edges begin to decrease.
We observe that both $n_3$ and $n_4$ increase with $\phi$ because the network becomes denser with increasing $\phi$. As the packing fraction $\phi$ increases, $n_3$ increases more rapidly than $n_4$.
The analysis of $n_l$ for $l>4$ is provided in Supplementary Information (SI).

%
%

The increase in $n_3$ (and $n_4$) with both $\phi$ and $\tilde{\sigma}$ appears to correlate with the rheology of the dense frictional suspensions, as reflected in $\eta_r$ (and normal stresses, as presented in~\cite{Singh_2018}). This correlation suggests that the mesoscale structure of the network, characterized by loop order, serves as a valuable order parameter that defines the macroscopic state of the material. The state of the material is reflected not only in the growth of frictional contacts but also in the evolving topological structure of the network as a function of stress and packing fraction.

\paragraph*{Effect of friction.}
Previous studies have shown that the shear rheology of dense suspensions is governed by three key parameters: packing fraction $\phi$, dimensionless shear stress $\tilde{\sigma}$, and the friction coefficient $\mu$~\cite{Mari_2014,Singh_2018,Singh_2020, Singh_2022}. After examining the effect of $(\phi,\tilde{\sigma})$ for $\mu=1$, we now investigate the rheology and mesoscale network behavior for varying $\mu$ by sweeping through shear stress in the range $\tilde{\sigma} \in \{0.1, 100\}$ at a constant $\phi=0.78$.

\begin{figure}[ht!]
    \centering
    \includegraphics[width=1\textwidth]{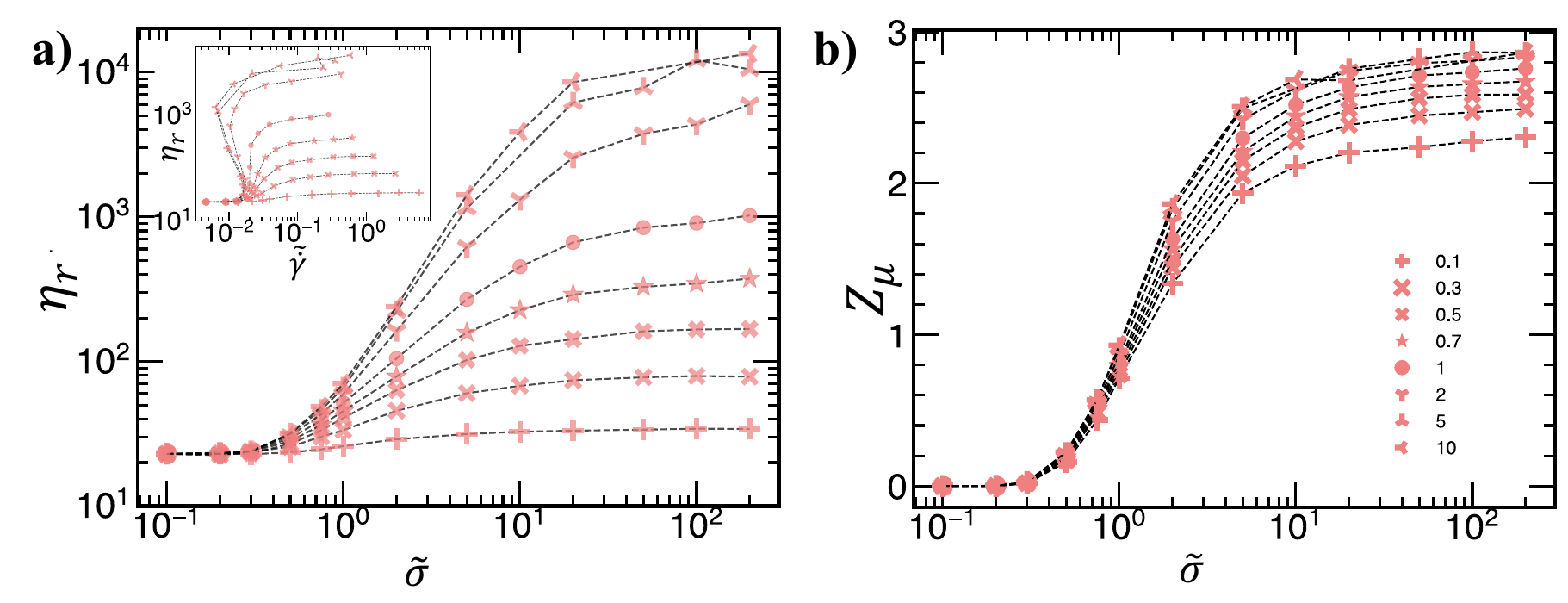}
    \caption{\textbf{Effect of friction.} (a) Relative viscosity $\eta_r$  and (b) average coordination number $Z_{\mu}$ as a function of scaled stress $\tilde{\sigma}$ for different friction $\mu$ at $\phi$=0.78. 
    The inset in (a) shows relative viscosity $\eta_r$ as a function of shear rate $\tilde{\dot{\gamma}}$ showing DST for $\mu \ge 1$.}
    \label{fig7-rheofixphi}
\end{figure}

Figure~\ref{fig7-rheofixphi} illustrates the influence of the static friction coefficient $\mu$ on the steady-state viscosity $\eta_r$ and frictional coordination number $Z_\mu$. In agreement with previous studies~\cite{Singh_2018, Mari_2014}, we observe that $\mu$ significantly impacts the viscosity $\eta_r$. Weak shear thickening is observed for $\mu=0.1$, whereas a discontinuous shear thickening (DST) is observed for $\mu \ge 1$. Because frictional contacts in our simulations are stress-activated, the suspension behaves as frictionless at $\sigma \ll \sigma_0$, making the viscosity independent of $\mu$. As the stress increases, frictional contacts are activated. Therefore, the extent of shear thickening is determined by $\phi_J^\mu-\phi$, where higher $\mu$ reduces $\phi_J^\mu$ making the shear thickening more severe.

Figure~\ref{fig7-rheofixphi}b shows $Z_\mu$ as a function of scaled stress $\tilde{\sigma}$.
As $\mu$ increases, the force chains that bear the load are stabilized, allowing more frictional contacts to withstand external stress. This results in a higher steady-state average $Z_\mu$ as a function of stress. This is also consistent with previous findings for dry granular materials~\cite{Singh_2013}.

\begin{figure}[ht!]
    \centering
    \includegraphics[width=1\textwidth]{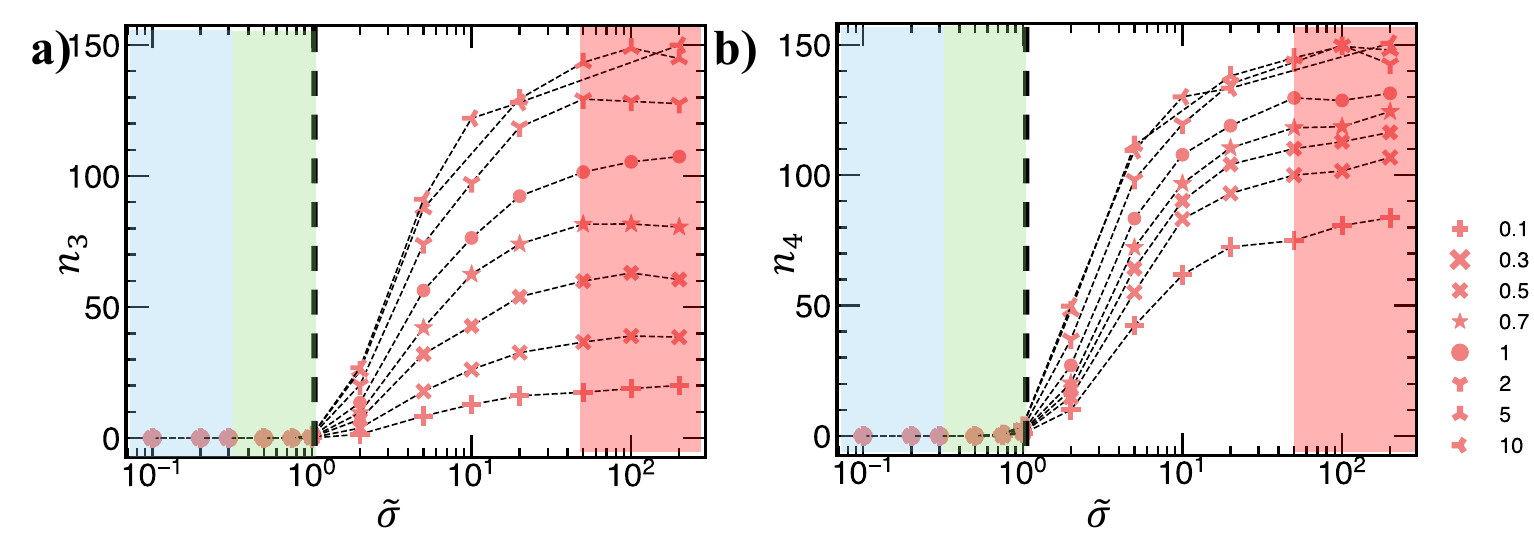}
    \caption{\textbf{Effect of friction on network topology.} Number of (a) third-order loops ($n_3$) and (b) forth-order loops ($n_4$) as a function of scaled stress $\tilde{\sigma}$  at packing fractions $\phi$ for different $\mu$. The color coding for the loop is the same as that in Fig. \ref{fig5-evon12}.}
    \label{fig8-evofixphi}
\end{figure}

To gain mesoscale insights into how friction affects the network topology, we use the loop-detection algorithm on simulation snapshots with different $\mu$. The results for non-closed components, that is, isolated and connected edges, are presented in SI (Fig.~S6), where we find the results to be consistent with previous analysis of variation in $\phi$ (Ref.~Fig.~\ref{fig5-evon12}).
Next, Fig.~\ref{fig8-evofixphi} shows the number of third- and fourth-order loops as a function of scaled stress $\tilde{\sigma}$ at a constant packing fraction $\phi=0.78$ for different values of $\mu$.
We find that both $n_3$ and $n_4$ increase as $\mu$ increases, but there is a striking difference in the extent of their increase as $\mu$ changes. For example, at $\tilde{\sigma}=100$, $n_3$ increases roughly 7.5 times as $\mu$ increases from 0.1 to 10; while $n_4$ roughly doubles. This observation highlights that changes in the number of loops are not just a reflection of the increase in $Z_\mu$, but that the network is topologically different as $\mu$ increases. 
%

\paragraph*{Connecting microscopic features with a macroscopic response via mesoscale features.}
Recent advances in the study of dense frictional suspensions suggest that the stress-induced frictional coordination number plays a crucial role in driving shear thickening behavior~\cite{Seto_2013a, Mari_2014, Wyart_2014, Ness_2016, Singh_2023}. This raises an intriguing question: can the extensive data collected across a wide range of parameter spaces—spanning packing fraction $\phi$ and applied stress $\tilde{\sigma}$ and friction coefficient $\mu$—be collapsed by simply using the stress-activated frictional coordination number $Z_\mu$?

\begin{figure}[ht!]
    \centering
    \includegraphics[width=1\textwidth]{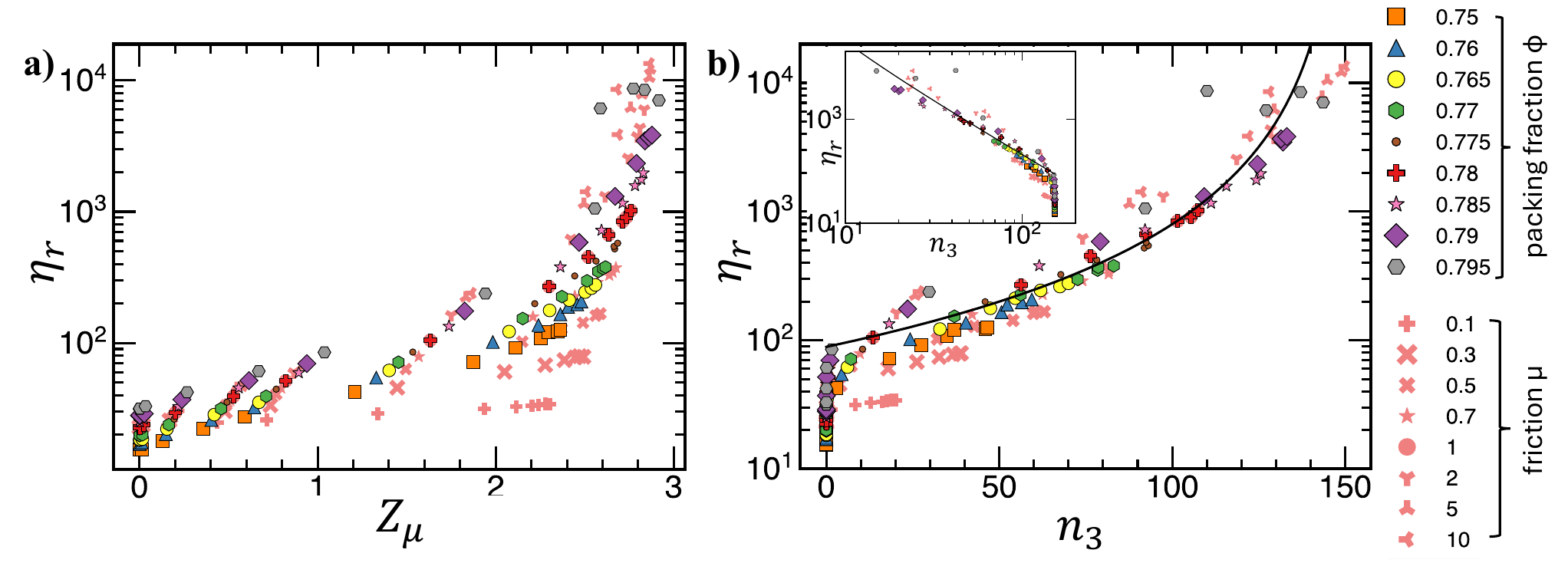}
    \caption{\textbf{Connecting rheology with mesoscale network topology.} Relative viscosity $\eta_r$ vs (a) frictional coordination number $Z_\mu$ and (b) number of third order loops $n_3$ for data with variable friction at $\phi=0.78$ and variable packing fraction for $\mu=1$. The solid line represents $\eta_r \propto (n_3^\mathrm{J} - n_3)^{-\alpha}$ with $n_3^\mathrm{J}=150$ and $\alpha=2$. (Inset) The same data on a log-log scale shows $\eta_r$ vs $(n_3^\mathrm{J} - n_3)$. The solid line shows the power law of $-2$.}
    \label{fig9-con}
\end{figure}

We explore this idea in Fig.~\ref{fig9-con}a by plotting the viscosity $\eta_r$ against the frictional coordination number $Z_\mu$ for all the data sets generated in this study. Contrary to our expectations, data collapse or a straightforward one-to-one relationship is not observed. Instead, the data suggest the presence of different families of $\eta_r(\phi,\sigma, Z_\mu)$ curves, which do not converge into a single master curve. In contrast, plotting viscosity $\eta_r$ as a function of the number of third-order loops $n_3$ yields a better collapse of the data, Fig.~\ref{fig9-con}b. Figure~\ref{fig9-con}b shows that the viscosity increases with $n_3$ and appears to diverge at a maximum $n_3$. The solid line shows $\eta_r \propto (n_3^\mathrm{J} - n_3)^{-\alpha}$ with $n_3^\mathrm{J}=150$ and $\alpha=2$. The higher $\phi$ and $\mu$ data sets, in which DST or non-monotonic flow-curves are observed, follow this scaling law, while data sets with low $(\phi,\tilde{\sigma},\mu)$ do not agree with the
proposed correlation. It is noteworthy that there are combinations of $(\phi,\sigma,\mu)$ that have similar $Z_\mu$ but different values of viscosity. Analyzing $\eta_r(n_3)$ separates these data sets into two families, which we uncover below.

From Fig.~\ref{fig9-con}b, we can observe that during shear thickening, the increase in viscosity follows a two-step process. The initial increase, roughly one order of magnitude, is primarily driven by an increase in the packing fraction $\phi$ and the activation of frictional contacts that do not necessarily form loops (only isolated and connected edges). At higher values of the parameter space $(\phi, \tilde{\sigma}, \mu)$, third-order loops, the smallest minimal rigid structures~\cite{Laman_1970}, begin to form, and their increase leads to a further increase in viscosity.
In other words, a smaller number of third-order loops (which arise from low $\phi$, $\tilde{\sigma}$, or $\mu$ values) results in lower viscosity $\eta_r$, whereas a larger number of these loops provides local rigidity to the suspension, thereby increasing $\eta_r$. The fact that the obtained $\eta_r(n_3)$ master curve is independent of the packing fraction, shear stress, and interparticle friction underscores the critical role of mesoscale structure as the bridge connecting microscale properties (such as particle friction) to macroscale rheology.
This correlation suggests that the viscosity $\eta_r$ is uniquely governed by the mesoscale topology, which we capture using the number of third-order loops. The observed diverging behavior implies that $\eta_r$ is regulated by $n_3$ or its distance from $n_3^{\mathrm{J}}$. It is important to note that different datasets corresponding to different $\mu$ values have varying intrinsic frictional jamming points $\phi_J^\mu$, but their corresponding $n_3^{\mathrm{J}}$ values remain similar. Finally, it is worth emphasizing that for a given combination of $(\phi, \tilde{\sigma}, \mu)$ that results in the same $n_3$ at the mesoscale, the system exhibits a comparable macroscale response, specifically yielding the same relative viscosity.

\section{Concluding Remarks}\label{Conc}
In this study, we focus on the topological features of the sub-network of frictional contacts in sheared dense suspensions since prior studies have shown that this sub-network drives strong shear-thickening~\cite{Mari_2014, Singh_2018, Gameiro_2020, Thomas_2018, Singh_2020}.
We show that strong shear thickening arises from the mesoscale structures in the frictional network, which we identified as ``loops.''
We identified that third-order loops show a one-to-one correlation with the viscosity, which is insensitive to change in $\phi,\tilde{\sigma}, \mu$ (given that stress is high enough, the coefficient of friction high enough ($\mu \gtrsim 0.3$), and the packing fraction analyzed is close to $\phi_J^\mu$). We find the existence of higher-order loops $l \le 8$, but none show a strong correlation with viscosity as presented by the third-order loops, being insensitive to the parameter space ($\phi,\tilde{\sigma},\mu$) explored.
At small stress, just at the onset of ST (and not DST), we find isolated and connected edges leading to an increase in viscosity. These structures do not contribute much to rigidity themselves but are needed to form third- and higher-order structures.
Notably, we have explored the number of loops, not their interconnectivity or percolation. In fact, Supplementary Videos show that the loops are highly heterogeneous and the third-order loops never percolate and yet seem to relate to the viscosity divergence~(Fig.~\ref{fig9-con}b).
 This highlights the subtlety of our findings that the increase in loops is not just a consequence of the increase in the coordination number but is also a product of the distinct topology as the system is sheared.

\paragraph*{Relationship with Previous Studies.} Our results highlight that a simple description of the network using the number of loops, particularly the number of third-order loops, $n_3$—serves as a mesoscale order parameter, providing a unique prediction of rheological behavior. Notably, the independence of our master curve from $\mu$ is a significant finding, especially given that previous studies using minimal rigidity models, such as the pebble game, did not yield robust results with varying $\mu$~\cite{Naald_2024}. In the structural rigidity theory, a 3-cycle, the third-order loop (here), has been hypothesized to be the smallest \textit{minimal} rigid structure in two dimensions. According to Lamans' theorem, triangles are the smallest isostatic structures that do not continuously deform under externally applied load. In contrast, 4-cycles (or 4th-order loop, here) continuously deform from one configuration to another while preserving lengths and connectivity~\cite{Laman_1970,Asimow_1978,Crapo_1979}. This could explain why $\eta_r(n_3)$ is insensitive to $\mu$, whereas $\eta_r(n_4)$ is not (Fig.~S7, SI).
In previous granular matter studies, even and odd order loops (or cycles) were also associated with allowing grains to roll and frustrated rotations, respectively~\cite{RIVIER20064505}. Other studies have also interpreted contact loops as stabilizing mesoscale features in the context of jamming/unjamming and shear-banding~\cite{Arevalo_2010,Walker_2010,Tordesillas_2010, Smart_2008}. Thus, it is no surprise that we find that the third-order (and not fourth-order)loops to correlate remarkably with the viscosity.

In this work, we focused on steady-state, strain-averaged mesoscale features and their relationship to rheology, leaving the exploration of temporal evolution and fluctuations for future studies. We only explored the mesoscale structure of the evolving contact network; future studies will relate the frustrated/free rotation of particles in an odd/even order loop, as shown in a dry granular context. We anticipate that our approach, which analyzes complex network structures in terms of loops, could offer valuable insights into the behavior of other amorphous materials, such as emulsions, colloidal gels, and foams.

%
\begin{acknowledgments}
A.S. acknowledges the Case Western Reserve University for start-up funding.
This work used the High-Performance Computing Resource in the Core Facility for Advanced Research Computing at Case Western Reserve University.
\end{acknowledgments}


%
%


\bibliography{dst}
\clearpage
\begin{widetext}
\begin{appendix}
    \section*{Supplementary Information to: \\
Topological insights into dense frictional suspension rheology: Third order loops drive discontinuous shear thickening}

   In this document, we provide details about (i) average stress contribution per loop order $l$, (ii) number of loops $n_l$ vs $l$, (iii) contact network evolution with strain, (iv) evolution of loop structure, (v) higher order loop $n_5$-$n_8$, (vi) isolated and connected edges for different $\mu$, (vii) viscosity as a function of $n_4$. These analyses complement the ones presented in the main text. 

\beginsupplement

\section{Loop Order Analysis}
In the main text, we presented a rationale to constrain the loop calculation to $3\le l \le 8$. Figure~\ref{figs1-asc} shows average stress carried by each loop order $l$ as a function of $\phi$ for $\tilde{\sigma}=100$ and $\mu=1$. We notice that isolated edges contribute minimally to stress, which is expected. It is also intuitive that the stress carried by each closed structure will increase with $l$. Note that the increase in stress contribution for $6 \le l \le 8$ is minimal, almost collapsing with connected edges. Hence, we do not calculate structures beyond the eighth order, ensuring we preserve the important topological insights but making the computation tractable.

\begin{figure}[ht!]
    \centering
    \includegraphics[width=1\textwidth]{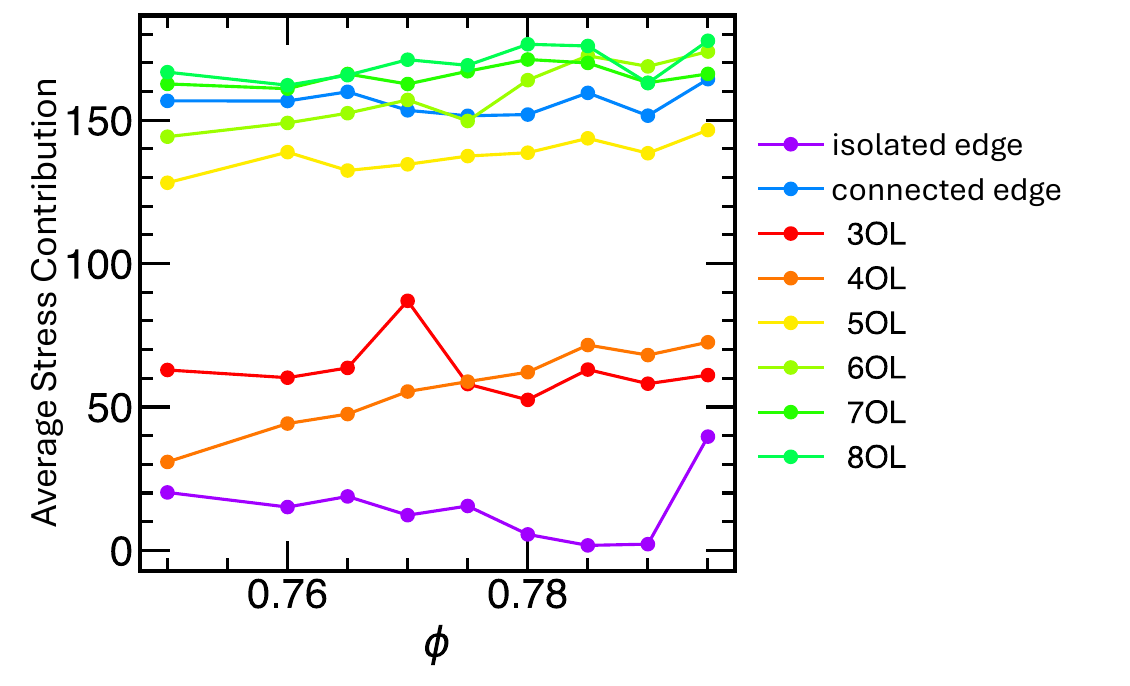}
    \caption{Average stress contribution at $\tilde{\sigma}$ = 100 for different type of loops at different packing fraction $\phi$ at $\mu$=1}
    \label{figs1-asc}
\end{figure}


\section{Loop Statistics}
Previous studies~\cite{Smart_2008} showed that the number of loops $n_l$ decrease with the order $l$. We confirm this finding in Fig.~\ref{figs2-loopstability} with a subtle difference. First our overall findings are consistent with Smart \& Ottino~\cite{Smart_2008} that the number of loops $n_l$ decrease with increasing $l$ for all values of $\phi,\tilde{\sigma},\mu$ considered here. Smart \& Ottino showed a decrease in third-order loops with increasing $\mu$; however we find the opposite trend. First, we highlight that the two systems (ours and one presented by Smart \& Ottino) are different; we performed simulation at constant volume, whereas Smart \& Ottino simulated constant pressure allowing the volume to fluctuate. In our system, increasing friction brings system closer to its frictional jamming point $\phi_J^\mu$ thus allowing more locally rigid structures (third-order loops) to form.


\begin{figure}[ht!]
    \centering
    \includegraphics[width=1\textwidth]{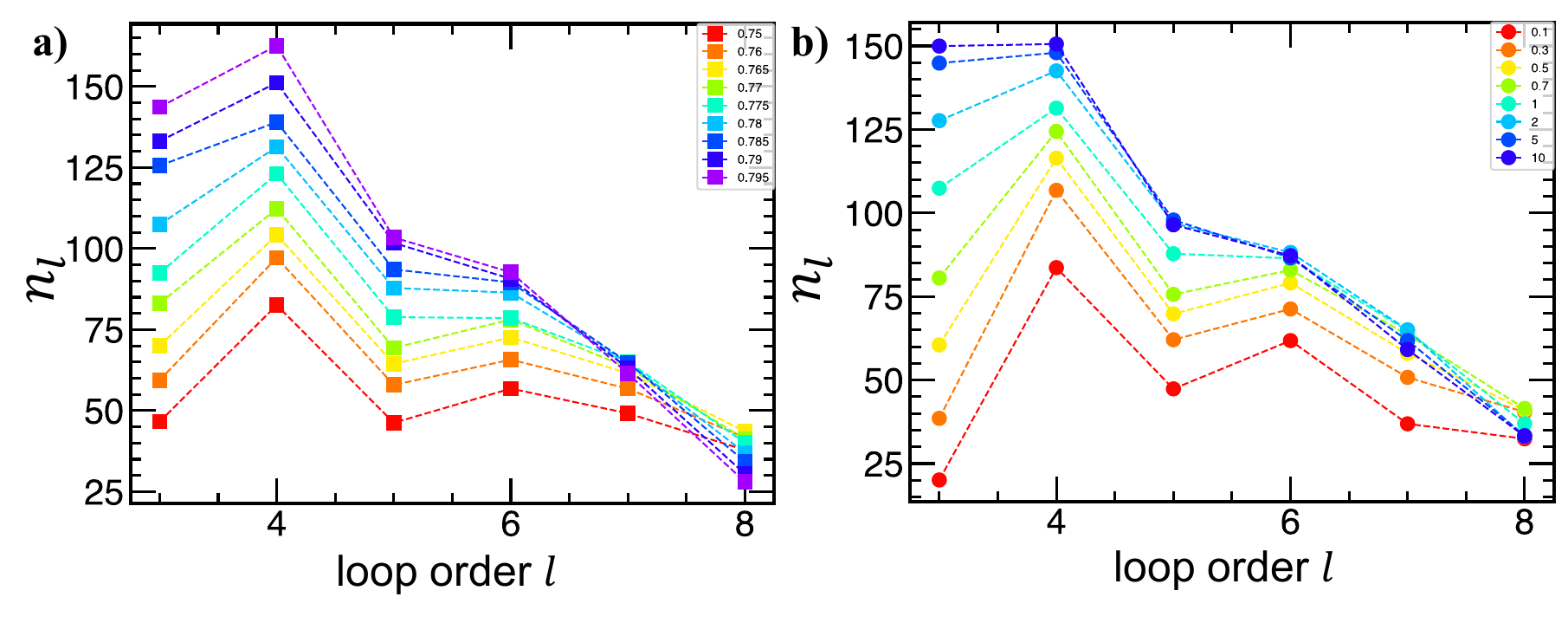}
    \caption{Number of loops as a function of loop order $n_l$ at $\tilde{\sigma}$ = 200 (a) varying packing fraction $\phi$ at friction $\mu$=1 and (b) varying friction $\mu$ at packing fraction $\phi$ = 0.78 }
    \label{figs2-loopstability}
\end{figure}


\section{Evolution of frictional contact network}
Figure 4 of the main text shows a representative snapshot for various values of $\tilde{\sigma}$. Figure~\ref{figs3-timenetwork} shows the frictional contact network for a few more strain values confirming that: (i) we indeed attain the steady state; (ii) at different snapshots (especially for $\tilde{\sigma}=1$) brings different particles into contacts highlighting the dynamic nature of our simulations.
\begin{figure}[ht!]
    \centering
    \includegraphics[width=1\textwidth]{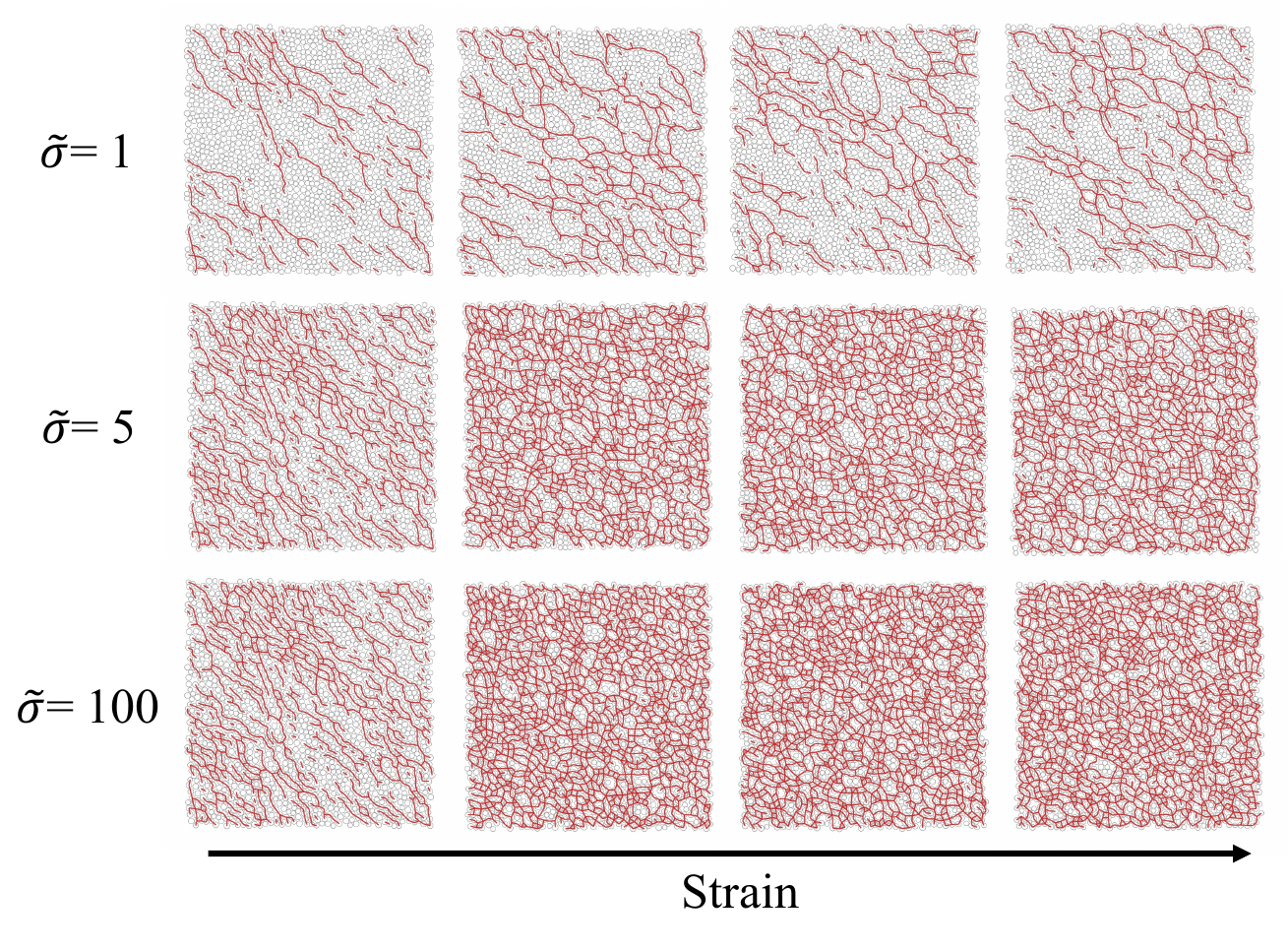}
    \caption{Strain evolution of frictional contact network at friction $\mu$ = 1 and packing fraction $\phi$ = 0.78 at three stress levels $\tilde{\sigma}$: 1, 5 and 100. Strain increases from left to right. The frictional contacts are shown in red.}
    \label{figs3-timenetwork}
\end{figure}

\section{Evolution of network topology}
Figure 4 of the main text shows a representative snapshot and corresponding loop structure for various values of $\tilde{\sigma}$. Figure~\ref{figs4-timeloop} shows the loop structure obtained from the frictional contact network for a few more strain values.

\begin{figure}[ht!]
    \centering
    \includegraphics[width=1\textwidth]{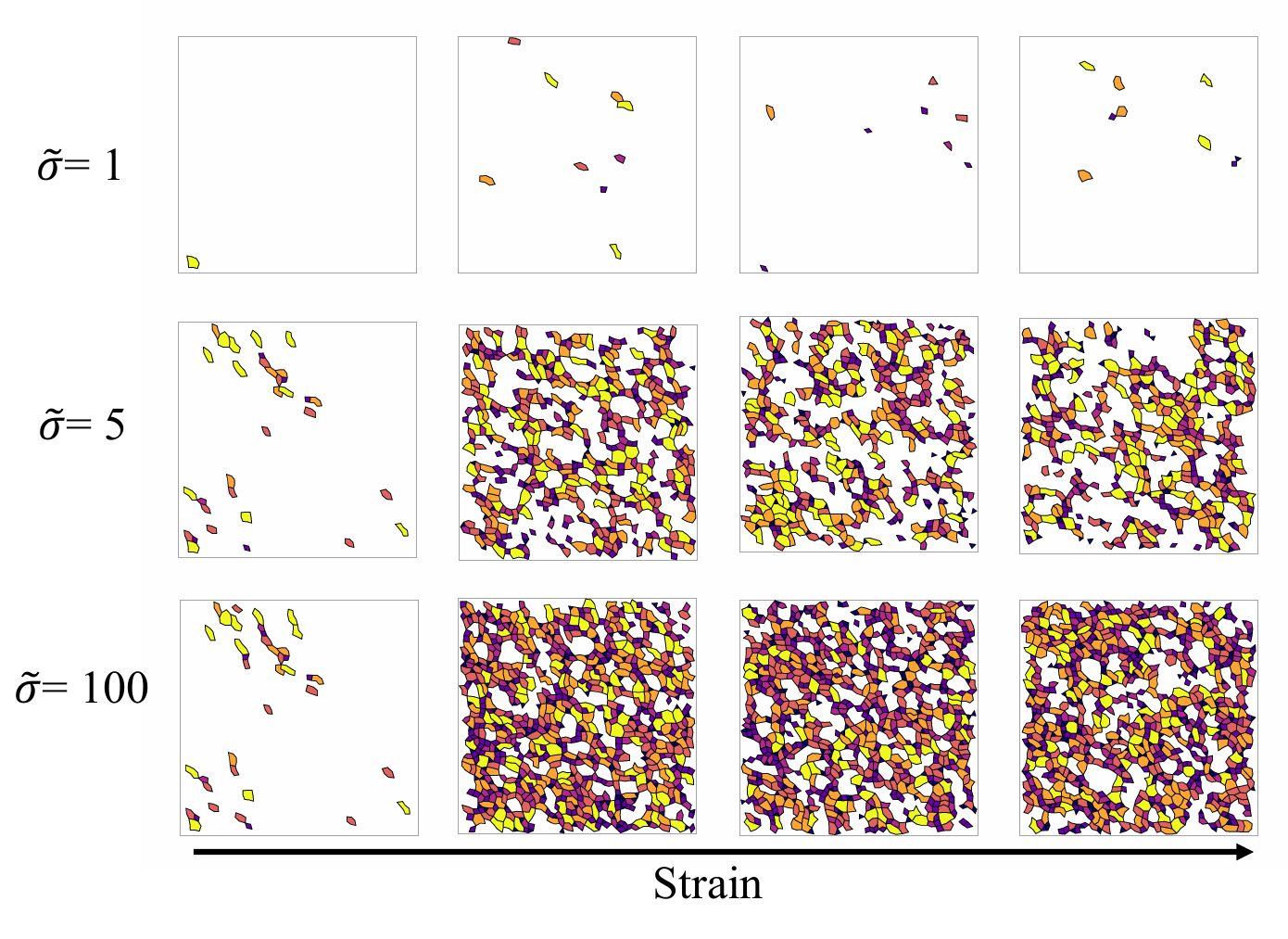}
    \caption{Time evolution of loops at friction $\mu$ = 1 and packing fraction $\phi$ = 0.78 at three stress levels $\tilde{\sigma}$: 1, 5. The color coding for the loop is the same as in Fig. 1 of the main text. Strain increases from left to right.}
    \label{figs4-timeloop}
\end{figure}

\section{Evolution of higher-order loops $n_5$-$n_8$}
In the main text, we only presented the isolated- and connected-edges along with third- and fourth-order loops. Figure~\ref{figs5-highorder} shows the evolution of fifth to eight-order loops as a function of $\tilde{\sigma}$ for various values of $\phi$ for $\mu=1$. We observe that the sensitivity to increase in $\phi$ decreases as the loop order increases. As an example, at a stress of 100, $n_5$ increases roughly 1.6 folds as volume fraction increases, $n_7$ shows barely any change with $\phi$ (at the same stress). $n_8$ is roughly insensitive to $\phi$ at higher stresses.

\begin{figure}[ht!]
    \centering
    \includegraphics[width=1\textwidth]{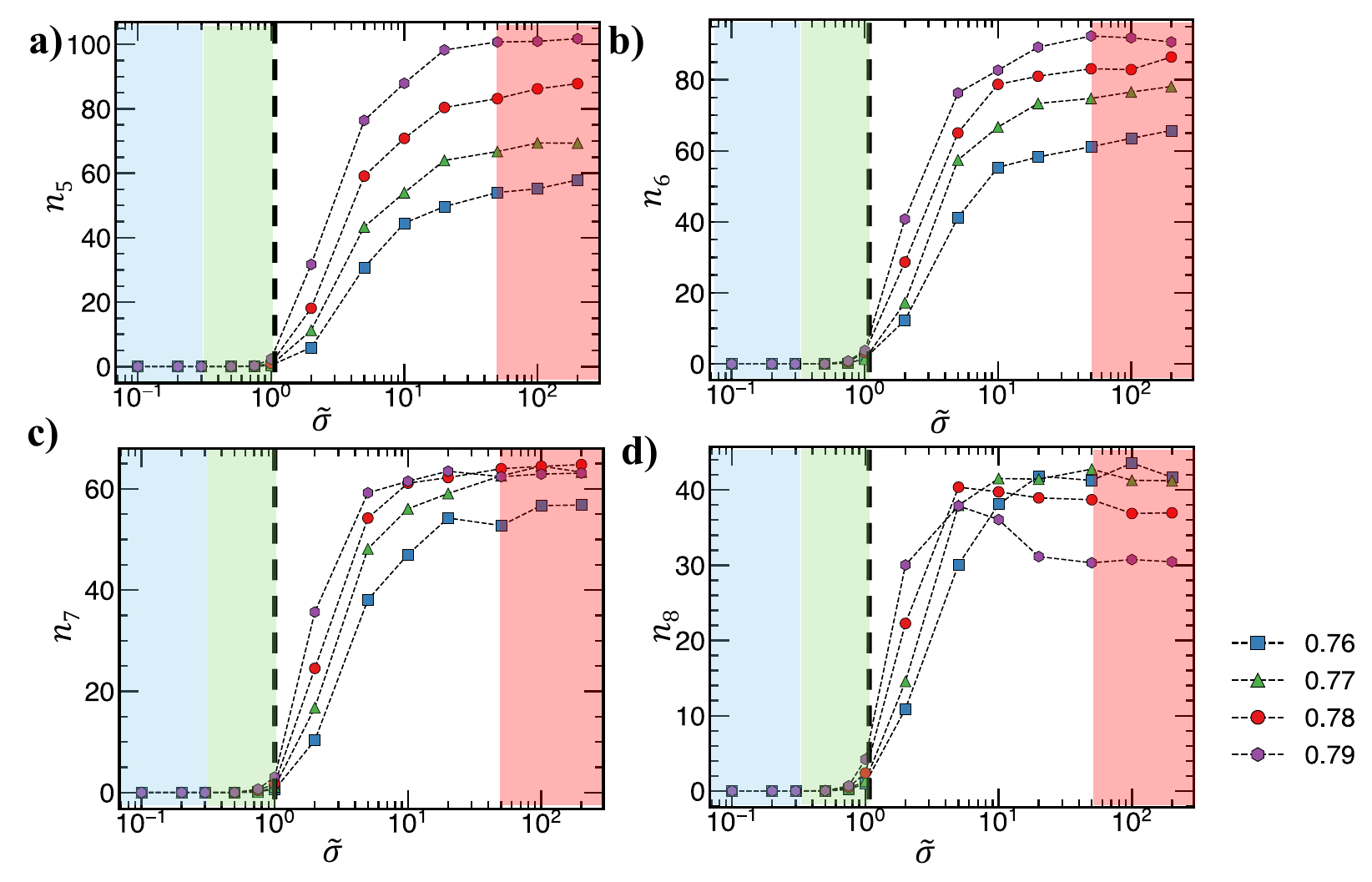}
    \caption{Number of (a) fifth order loop ($n_5$), (b) sixth order loop ($n_6$), (c) seventh order loop ($n_7$) and (d) eighth order loop ($n_8$) as a function of scaled stress $\tilde{\sigma}$ at $\mu$=1 for different packing fractions $\phi$. The color coding for the loop is the same as in Fig.5. of the main text.}
    \label{figs5-highorder}
\end{figure}

\section{Evolution of isolated and connected-edges for different $\mu$}
In the main text, we only presented the evolution of third- and fourth-order loops for simulations with different $\mu$. Figure~\ref{figs6-edge_mu} shows the evolution of isolated- and connected-edges as a function of stress for different values of $\mu$. We observe that increasing $\mu$ leads to a decrease in isolated and connected edges at higher stresses.

\begin{figure}[ht!]
    \centering
    \includegraphics[width=1\textwidth]{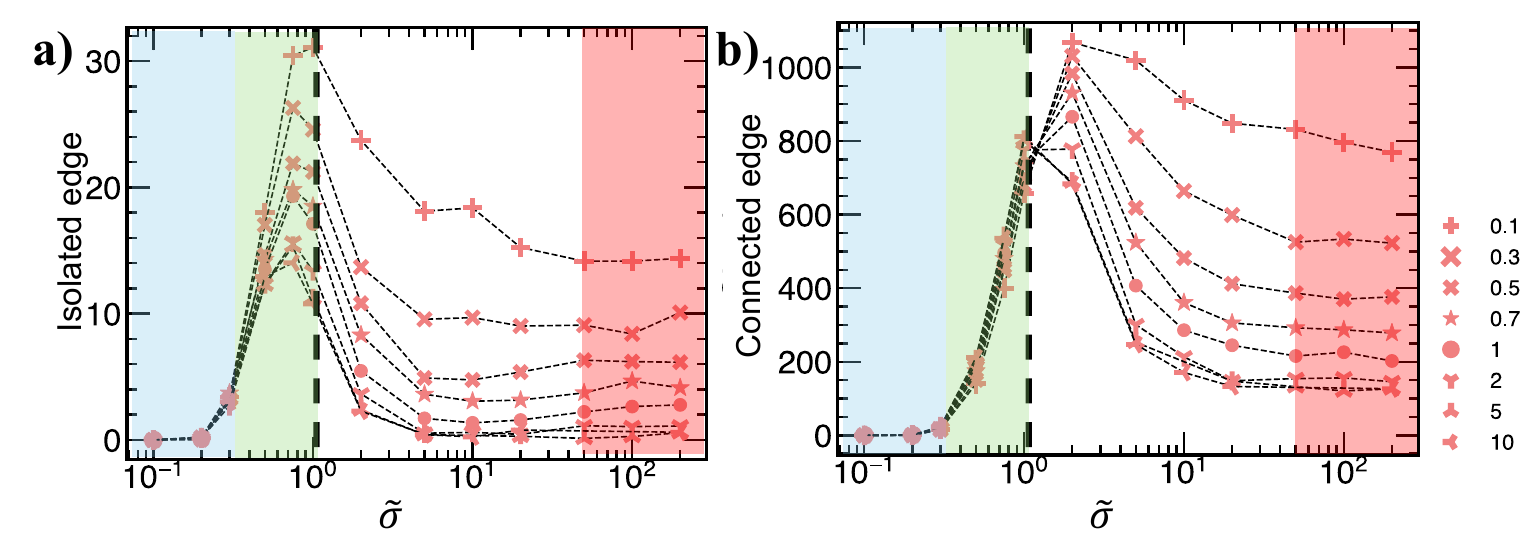}
    \caption{Number of (a) isolated edge and (b) connected edgeas a function of scaled stress $\tilde{\sigma}$ at $\phi$=0.78 for different friction $\mu$. The color coding for the loop is the same as in Fig.5. of the main text.}
    \label{figs6-edge_mu}
\end{figure}

\section{Viscosity as a function of number of fourth order loops}
Figure~\ref{figs7-evofixmu} shows viscosity plotted as a function of the number of fourth-order loops $n_4$ for all the simulation data presented in this study. We do not observe the data has a reasonable collapse as analyzed for a constant $\mu$; while data for different friction, falls off the collapse (or master curve). This highlights the significance of third-order loops as compared to the fourth-order loops.
\begin{figure}[ht!]
    \centering
    \includegraphics[width=1\textwidth]{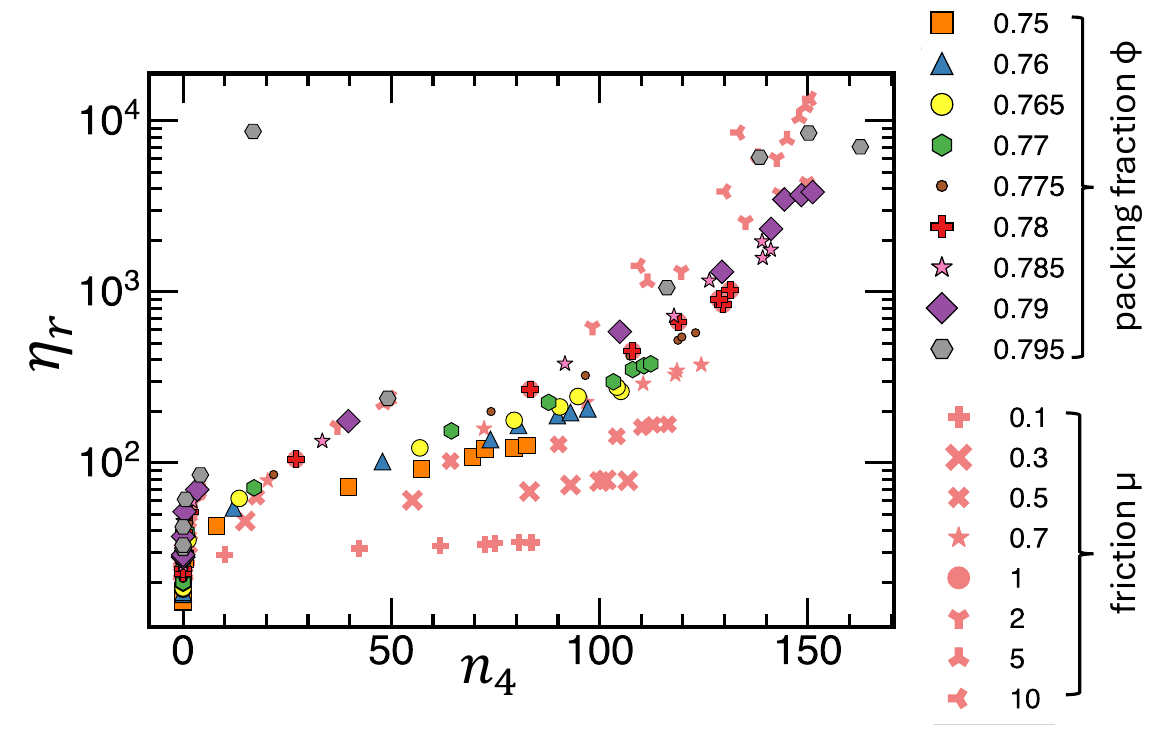}
    \caption{Master curve of relative viscosity $\eta_r$ as a function of forth loop order $n_4$}
    \label{figs7-evofixmu}
\end{figure}

\section{Movie}
The movie shows the evolution of network topology (loops) with packing fraction and stress in Fig. b-d (and f-h) of the main text.

\clearpage

\end{appendix}
\end{widetext}
\end{document}